# Direct Graphene Growth on Insulator

Gunther Lippert[*,1], Jarek Dabrowski[1], Max C. Lemme[2,3], Charles M. Marcus[2], Olaf Seifarth[1], Grzegorz Lupina[1]

[1] Innovations for High Performance microelectronics, Im Technologiepark 25, 15236 Frankfurt (Oder), Germany
[2] Department of Physics, Harvard University, Cambridge, MA 02138, USA
[3] KTH Royal Institute of Technology, 16640 Kista, Sweden

* Corresponding author: lippert@ihp-microelectronics.com, Phone: +49 335 5625 727, Fax: +49 335 5625 681

**Abstract**

Fabrication of graphene devices is often hindered by incompatibility between the silicon technology and the methods of graphene growth. Exfoliation from graphite yields excellent films but is good mainly for research. Graphene grown on metal has a technological potential but requires mechanical transfer. Growth by SiC decomposition requires a temperature budget exceeding the technological limits. These issues could be circumvented by growing graphene directly on insulator, implying Van der Waals growth. During growth, the insulator acts as a support defining the growth plane. In the device, it insulates graphene from the Si substrate. We demonstrate planar growth of graphene on mica surface. This was achieved by molecular beam deposition above 600°C. High resolution Raman scans illustrate the effect of growth parameters and substrate topography on the film perfection. Ab initio calculations suggest a growth model. Data analysis highlights the competition between nucleation at surface steps and flat surface. As a proof of concept, we show the evidence of electric field effect in a transistor with a directly grown channel.

**Introduction**

Graphene is a 2D crystal of sp²-bonded carbon in a honeycomb lattice. One of the outstanding properties of graphene with one or two layers is the high charge carrier mobility and high saturation velocity, which could be employed to revolutionize high-frequency microelectronic [1]. Bi-layer graphene is of particular interest, as its band gap can be controlled [2]. Other potential applications of graphene include optoelectronics [3] and chemical sensors [4]. Graphene can be exfoliated from graphite, grown on metal and transferred to the target wafer [5] or grown epitaxially on SiC [6]. Even though the two latter schemes have a potential for mass production, there are technological short-comings. Growth on metal requires subsequent mechanical transfer and hence is not necessarily compatible with existing silicon processing technology. SiC wafers are limited in size and are quite expensive, and while it is possible to grow graphene epitaxially from SiC seeds deposited on Si, the required temperatures for either option are beyond typical silicon processes (>1000°C) [7].

Graphene deposition using plasma-enhanced chemical vapor deposition (PECVD) can meet the technology requirements, but PECVD films grown on insulators are so far either thick (thicker than six layers), strongly distorted, or do not grow parallel to the substrate [8]. We report on direct growth of graphene by molecular beam epitaxy (MBE) of C on an insulating substrate. For this study, mica (a layered alumina silicate) is selected as the substrate. Mica surfaces belong to the flattest known: it is possible to prepare mica samples free of steps on areas of macroscopic size. Layers of muscovite mica consist of three monatomic sub-layers: aluminum oxide is sandwiched between silicon dioxide. The layers are only weakly bonded, so that flat mica flakes with widely spaced steps can be obtained. When the step-step distance was in the micrometer range, formation of graphene is observed, with thicknesses down to two layers and with preference for even number of layers. With support of ab initio calculations it is argued that high-quality graphene nucleates at surface steps and grows by step flow. This conclusion should be general for Van der Waals growth of graphene on in-sulators. Together with the demonstration of successful growth on mica, it opens the way to silicon-compatible growth of device-size, device-quality graphene, including bi-layer films.

**Experimental**

Carbon was deposited in a molecular beam epitaxy (MBE) ultra-high vacuum (UHV) system from DCA on samples freshly cleaved from mica with trigonal (3T) crystal structure; the cleaved surface is thus (0001) and terminated with K [9]. The growth rate was of the order of a monolayer per minute. The source (high-purity pyrolytic carbon) emitted mostly atoms with a weak admixture of dimers [10]. It was placed 35 cm away from the sample. The substrate temperature was varied between 20°C and 1000°C and the growth time between 50 s and 600 s. The growth pressure was in the range of $10^{-7}$ mbar. A schematic of the proposed growth process is shown in Figure 1. Film thickness and quality was observed ex-situ by µ-Raman spectroscopy with a spatial resolution of 0.4 µm and a spectral resolution better than 2 $cm^{-1}$.

Electrical contacts to the graphene were defined by conventional electron beam lithography using polymethylmethacrylate (PMMA) resist, followed by evaporation of Ti and Au (5 nm/40 nm). The gate insulator was grown by atomic layer deposition (ALD); deposition of a functionalization layer based on $N_2O$ was followed in-situ by ALD of 20 nm of aluminium oxide ($Al_2O_3$) using a trimethylaluminum precursor [11]. Finally, the gate electrodes were defined by electron beam lithography and deposited by Ti / Au (5 nm/40 nm) thermal evaporation. The devices were measured in vacuum ($5\times10^{-3}$ mbar).

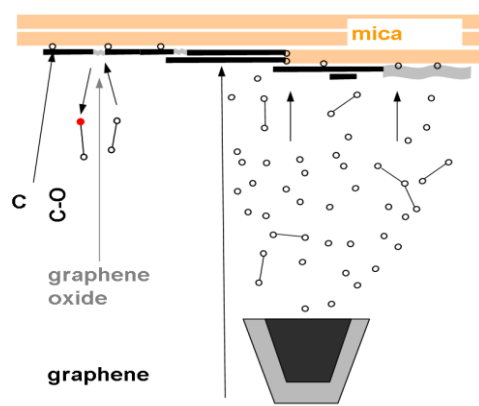

**Figure 1**: Energetic carbon atoms impinge onto the mica. Few-layer graphene nucleates at step edges and at chemisorbed carbon atoms (black circles). At temperatures above 600°C, nano-sized graphene flakes can slide freely to coalesce into a graphene film.

We also performed ab initio density functional theory (DFT) calculations for total energies, atomic structures, energy barriers, and molecular dynamics (in the range of picoseconds), using Quantum Espresso [12]. We carried out calculations for C atoms and small C clusters (dimers, hexagons, and nano-flakes with up to 42 C atoms). A simplified mica muscovite model was used: a single sheet with a central $Al_2O_3$ layer, two layers of hexagonal $SiO_2$, and surface K atoms. We considered chemisorption of C on mica surfaces, whereby a step was approximated by a $KAl_2Si_6O_{10}(OH)_{11}$ molecule either placed on top of 3×3 surface of the model mica, or suspended in vacuum.

## Results and Discussion

### Ab initio Calculations

Carbon clusters are physisorbed, unless attached to a carbon atom already chemisorbed on the surface or at a step (Figure 2). The physisorption energy is 0.18 eV/atom for $C_6$ and 0.05 eV/atom for $C_{24}$, i.e., clearly larger than the contribution from pure London dispersion force [13]. This shows that at least for smaller flakes adhesion is dominated by the interaction between permanent multipoles in mica and multipoles induced in the molecule (Van der Waals – Debye force).

However, individual C atoms can react with mica surface. First, they can reduce it by combining with the surface oxygen [14] and escaping as CO. Second, after desorption of CO, another carbon atom can fill the surface oxygen vacancy, forming substitutional surface carbon, $C_O$.

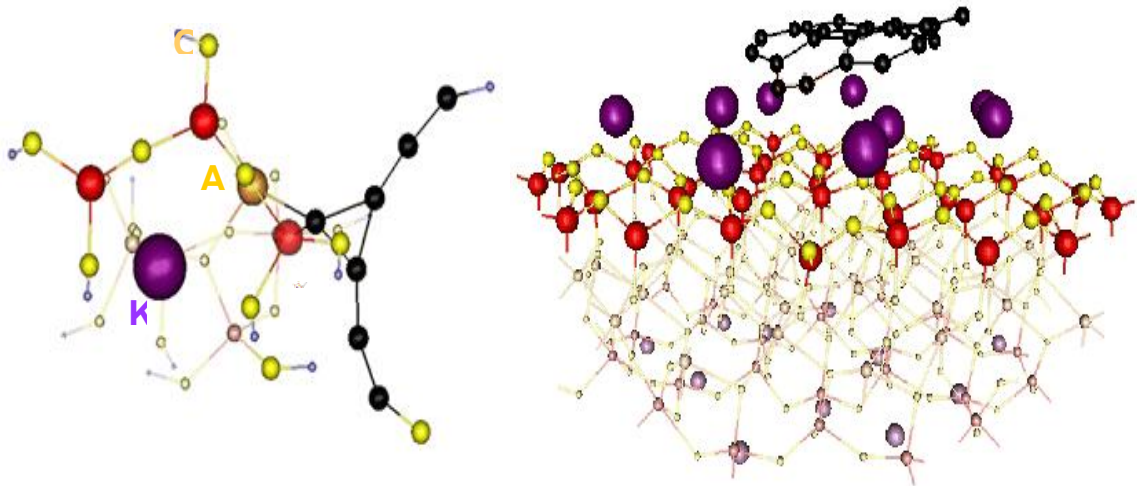

**Figure 2:** Ab initio calculations for C on mica. (Left) Hypothetical seed: C string grows at a surface step. (Right) Graphene molecule slides with no diffusion barrier on the surface to coalesce with other flakes (a frame from molecular dynamics [15]).

Molecular dynamics at 700°C shows that molecules even as big as $C_{24}$, if not attached to mica (e.g., at $C_O$), can slide on the surface [15] with thermal speed, i.e., with no barrier. When the molecule is attached at $C_O$, it can still migrate, but now with a barrier: the bonds with the $C_O$ must be passed from one C atom to another. Each docking point contributes around 1 eV to the barrier if the bonding is to C atoms far from the graphene edge. Consequently, already three such anchors may suffice to immobilize the affected molecule at 700°C. If the bonding is to the graphene edge, the barrier is expected to be also about 1 eV for diffusion along the edge, but for a complete detachment from it may exceed 5 eV ($C_O$-graphene bonding energy). Furthermore, we find that oxygen atoms tend to segregate to the growth front and that their presence facilitates the dissolution of graphene-substrate bonds.

**Growth at the Surface**

The deposition of carbon atoms on mica results produces islands of graphene. Figure 3 shows an optical micrograph of graphene grown on a flat area of mica at 900°C, taken with circularly polarized light. Four different regions in brightness contrast can be distinguished. Raman spectra taken in each of these areas are shown in the right panel of Figure 3. They are qualitatively different in each of the regions.

Judging from the Raman spectra (2D shape and 2D/G ratio), from the optical contrast in the optical microscope images, and from the correlation between the two, we assign the four regions labelled from 1 to 4 in Figure 3 to: (1) bi-layer graphene, (2) graphene with four carbon layers, (3) graphene with approximately six carbon layers [16,17], and (4) thicker graphene. We note that exact determination of the number of layers is hindered here by the fact that the interaction of grown graphene with the substrate modifies the line shape to a certain extent. This interaction seems to contribute also to the distortion peat at ~1350cm$^{-1}$. The small D/G peak and the G line width indicate that the quality of graphene is rather high.

These results, together with the ab initio data presented above, allow one to conclude how graphene is formed. The growth starts from seeds; on the flat surface, these may be $C_O$ defects mentioned above. The flakes become immobilized when they coalesce. Furthermore, oxygen helps to chemically bind the misaligned islands and thus becomes stabilized. This stabilization is to be expected also on the basis of ab initio thermodynamics of oxidation of carbon vacancy defects in graphene [18] In addition, oxygen-terminated cracks may be initiated at boundaries between misaligned islands and extend into larger flakes [19]. Particularly good graphene flakes are expected to be nu-cleated at step edges, because in this way growth by step flow can be achieved and the small, initially oxygen-interconnected islands may be incorporated into a relatively defect-free film. The majority of the oxygen atoms are then segregated to the growth front, returned to the substrate, or they escape in molecular form to the vacuum.

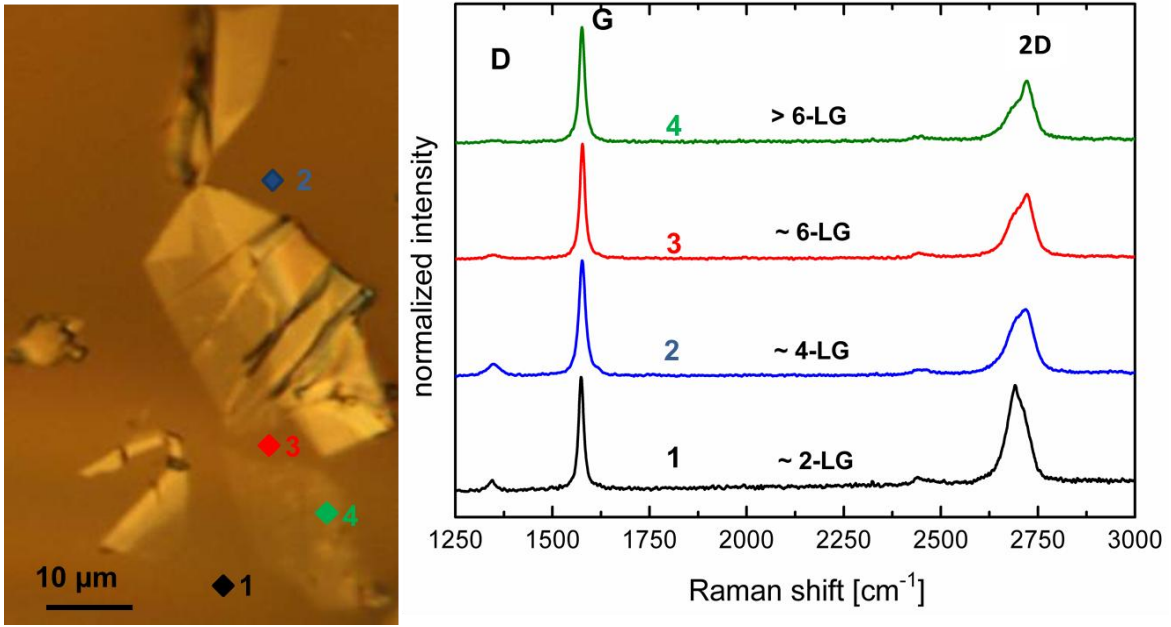

**Figure 3:** Few-layer graphene grown at 900°C: (left) as seen in optical microscope and (right) Raman spectra.

**Growth at Imperfections**

Figure 4 shows an area of disordered surface of mica (left) immediately after the start of growth. Raman mapping of this area indicates a decoration of step edges with graphene (right), as expected for a step flow driven growth. This indicates that steps act as seeds for the generation of good graphene. We note that growth by step flow is the preferred growth mode, because defect density should then decrease by the coalescence of flakes in a manner resembling the Oswald growth mechanism: the growing front acts as the edge of a huge island.

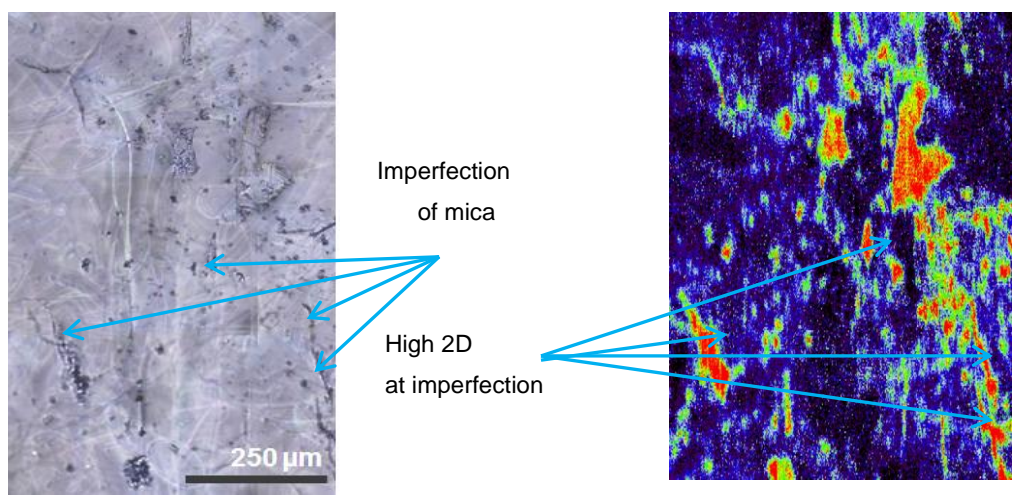

**Figure 4:** (Left) Mica step edges produce an optical contrast (Right) These sites are decorated by graphene, as visualized by Raman map of the 2D intensity (blue means low, red means high).

Step flow can explain the preference for films with an even number of layers. A muscovite mono-step (height of 0.7-0.8 nm) contains two natural nucleation sites for carbon layers. These are the oxygen bridges which, due to topology of the bond network of mica, must be broken at the step edge in the upper and lower surface of the topmost of the alumina silicate layers forming the layered (alumina silicate layers separated by potassium) mica structure. Graphene bi-layer can be also attached to the step through chemical bonds with carbon atoms substituting some of the oxygen atoms in the oxygen bridges interconnecting the mid sub-layer with the upper and lower sub-layer, but the estimated stability of such a configuration is substantially lower than the former two, particularly in the presence of oxygen. (Figure 2, left, illustrates a graphene chain attached to a substitutional carbon at a step kink).

This even-layered pattern is not necessarily a general feature. The multiple occurrences may be associated with particular properties of muscovite steps. Nevertheless, it may facilitate controlled growth of bi-layer graphene. Bi-layer graphene is of interest for applications, because its energy gap can be controlled by an electric field [2].

## Observation of the Electric Field Effect

Finally, we measured electrical transport through graphene field effect transistors (Figures 5 and 6) fabricated from few-layer graphene sheets grown by MBE. We applied a drain bias of 0.25 mV and measured the drain current while sweeping the gate voltage from -10 to 10 V.

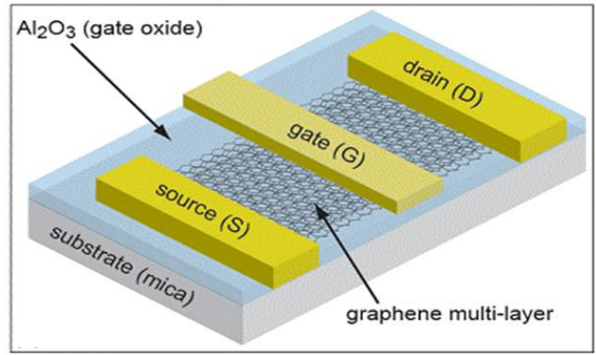

**Figure 5:** Schematic of a graphene field effect transistor grown on mica.

Figure 6a shows the device and 6b shows the resulting dependence of drain current on gate voltage. A clear response to the gate voltage can be observed. We attribute the absence of the typical charge neutrality point and the large noise to unintentional background p-doping during growth and processing. The drain current modulation is moderate, as expected for a multi-layer graphene device due to screening effects [21]. Nonetheless, the clear response of the drain current to the gate voltage demonstrates that transistors made from graphene films grown on an insulator exhibit the usual electric field effect.

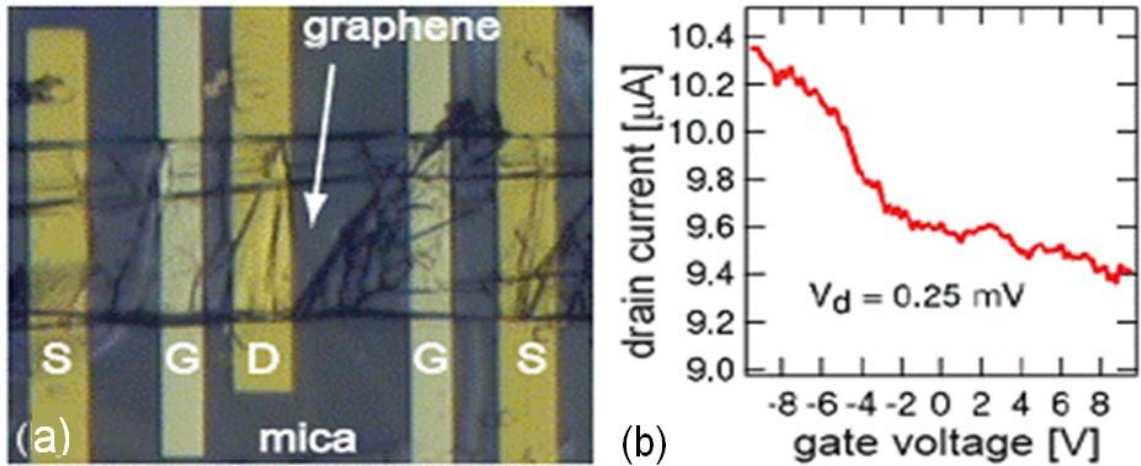

**Figure 6:** (a) Optical image of a graphene FET on mica.(b) Electrical data. Drain current was measured while sweeping the gate voltage ad drain bias of 0.25 mV. The resulting Id-Vg characteristic is typical for multi-layer graphene transistors.

## Summary and Conclusions

We experimentally observed and theoretically analysed the process of graphene growth on a Van der Waals substrate (mica).

Carbon atoms impinge onto the substrate surface. Few-layer graphene may nucleate at step edges and at chemisorbed carbon atoms. Due to the presence of efficient nucleation sites at steps, flakes with thinnest regions consisting of multiples of bi-layer graphene dominate. Graphene can also nucleate on a flat surface, where it becomes readily immobilized by carbon atoms, which in the initial stage of C deposition substituted some of the surface oxygen atoms. If a nano-sized graphene molecule immobilized on a flat surface is liberated (by thermal vibrations or by the action of oxygen in the rest gas), it can slide freely on the surface of graphene until it reaches a step of mica or of a terrace of graphene. This leads to a growth of high-quality graphene layer. Otherwise, the immobilized molecules coalesce and the quality is diminished by the defects produced at the boundaries of the coalesced islands. On the other hand, macroscopic flakes of graphene nucleated at surface steps and formed by step flow have defects mostly due to incomplete consumption of smaller islands. Graphene films bear a signature of their step origin: their thickness is predominantly a multiple of two layers.

We find that the growth temperature of 900°C is high enough for small carbon clusters to be sufficiently mobile, and low enough to prevent degradation of the substrate.

The growth mode proposed here to take place after the nucleation on flat surface may also be compatible with the recent report of CVD growth on $SiO_2$ [20]. Nano-sized graphene was observed to nucleate on the surface (catalytic action of nickel over-layer was then used to increase the size of these tiny islands).

Among the challenges ahead, the most obvious are related to the selection of a silicon-compatible insulator, to the deposition of seeds, and to the efficient suppression of heterogeneous growth. In addition, atomic-scale defects produced during growth must be understood and controlled. If these challenges are addressed, our approach outlines a realistic route to integrate (bi-layer) graphene into silicon process technology, and to fully exploit the potential of graphene as an electronic material.

**Acknowledgements**

We thank Ioan Costina, Wolfgang Mehr, Thomas Schroeder, Peter Zaumseil, Marvin Zöllner (all IHP) for experimental support and discussions. Ab initio calculations were performed in the Supercomputing Centre (JSC), Jülich, Germany (grant hfo06).